\documentclass[conference]{IEEEtran}
\IEEEoverridecommandlockouts
\usepackage{graphicx}

\usepackage{cite}
\usepackage{amsmath,amssymb,amsfonts}
\usepackage{algorithmic}
\usepackage{multirow}
\usepackage[none]{hyphenat}
\usepackage{float}
\usepackage{xcolor}
\usepackage{balance}
\usepackage[normalem]{ulem}
\usepackage{threeparttable}
\usepackage{pifont}

\usepackage[colorlinks=true,linkcolor=blue, citecolor=magenta]{hyperref}%

\usepackage[subtle]{savetrees}

\newcommand{\cmark}{\ding{51}}%
\newcommand{\xmark}{\ding{55}}%

  \DeclareUnicodeCharacter{202F}{FIX ME!!!!}

\def\BibTeX{{\rm B\kern-.05em{\sc i\kern-.025em b}\kern-.08em
    T\kern-.1667em\lower.7ex\hbox{E}\kern-.125emX}}
\begin{document}

\title{Non-Hermitian Physics-Inspired Voltage-Controlled Oscillators with Resistive Tuning}

\author{\IEEEauthorblockN{ Weidong Cao\textsuperscript{a}, Hua Wang\textsuperscript{b}, and Xuan Zhang\textsuperscript{a}}
\textsuperscript{a}Washington University in St. Louis, Saint Louis, MO, USA; \textsuperscript{b}ETH Zürich, Zürich, Switzerland  \\
}

\maketitle

\begin{abstract}
This paper presents a non-Hermitian physics-inspired voltage-controlled oscillator (VCO) topology, which is termed parity-time-symmetric topology.
The VCO consists of two coupled inductor-capacitor (LC) cores with a balanced gain and loss profile.
Due to the interplay between the gain/loss and their coupling, an extra degree of freedom is enabled via resistive tuning, which can enhance the frequency tuning range (FTR) beyond the bounds of conventional capacitive or inductive tuning.
A silicon prototype is implemented in a standard $130$ nm bulk CMOS process with a core area of  $0.15\text{mm}^2$.
Experimental results show that it achieves a $3.1\times$ FTR improvement and $30\%$ phase noise reduction of the baseline VCO with the same amount of capacitive tuning ability.

\end{abstract}


\section{Introduction}
\label{sec:intro}
Inductor-capacitor (LC) voltage-controlled oscillators (VCOs) are the key building blocks in many communication systems~\cite{fangxu,naiwen,caodfe}.
Such local oscillators need to cover a wide frequency tuning range (FTR) and maintain good phase noise (PN) performance. 
Existing methods mainly rely on aggressive inductive tuning (e.g., negative inductance/switched-inductors)~\cite{TCAS_I_12, VLSI_10_IND} or capacitive tuning (e.g., varactors/switched-capacitors)~\cite{JSSC_13} to reach a wide FTR. 
However, they often suffer from degenerated PN performance due to the excessive noise sources introduced by the tuning mechanisms. 
Some techniques have been proposed to tackle the stringent design trade-off between FTR and PN. 
For example, multi-core and multi-mode VCOs implemented with two or more coupled LC cores can reduce PN and extend the FTR, but bear the cost of excessive design complexity in VCO cores~\cite{ISSCC_19, JSSC_17}.
On the other hand, leveraging multiple separated LC VCOs can extend the FTR, but induces large area overhead and high multiplexing complexity~\cite{hai_kun}.
Therefore, developing more effective approaches to improving the FTR without hurting the PN performance is highly desirable.

Toward the goal, a gain-loss coupled dual-core VCO topology based on non-Hermitian physics is shown in this paper, which is originally proposed in our prior work~\cite{nn}.
From the perspective of non-Hermitian physics, the eigenfrequency of a physical system built upon two coupled units with a balanced gain and loss distribution can evolve in a wide range by tuning the gain/loss contrast~\cite{NM,nn}. 
Inspired by this physical principle, we build this gain-loss coupled dual-core VCO.
It not only inherits the superior PN performance from the coupled structure of multi-core VCOs but also achieves an enlarged FTR beyond the bounds of conventional capacitive/inductive tuning with extra resistive tuning.
The proposed topology can be combined with existing optimization approaches to further advance VCO performance.
Section~\ref{sec:bg} introduces the theoretical model of non-Hermitian quantum physical systems composed of two coupled units with a balanced gain and loss profile.
Circuit design and analysis are presented in Section~\ref{sec:imple}.
Comprehensive measurement results are shown in Section~\ref{sec:results} before the conclusion in Section~\ref{sec:conclu}.
\section{Background}
\label{sec:bg}

\begin{figure}[!t]
\centering
\includegraphics[width=1.0\linewidth]{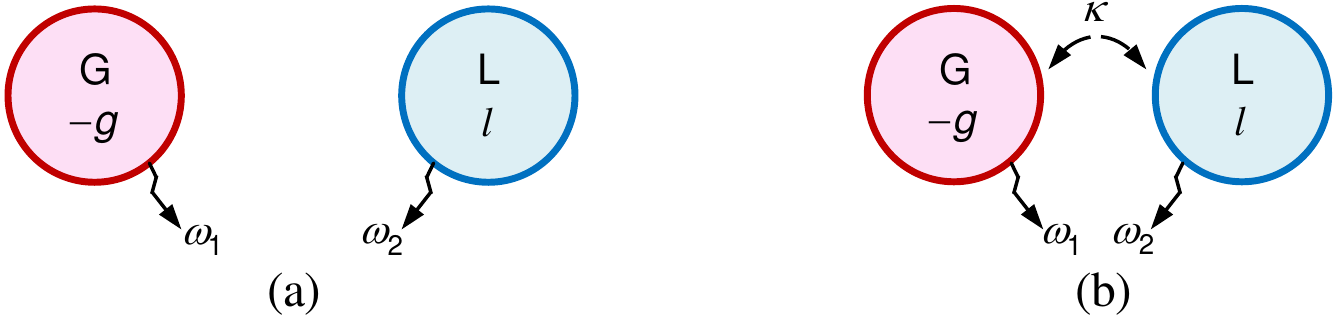}
\caption{(a) Illustration of an open non-Hermitian quantum system that can be modeled as a gain (G) or a loss (L) unit. (b) Illustration of a non-Hermitian quantum system with a coupled gain and loss profile. $\omega_{1,2}$ is the resonant frequency of the unit. $-g$ and $l$ represents gain and loss amount respectively.}
\label{fig:intro}
\end{figure}

\subsection{Non-Hermitian Quantum Mechanics}
\label{sec: pt-symmetry}
In quantum mechanics, an open system can be generally modeled as a gain (or loss) unit with a resonant frequency as shown in Fig.~\ref{fig:intro}(a).
Such systems are described by non-Hermitian Hamiltonians which preserve complex eigenvalues, i.e., $\omega_{1}-ig$ or $\omega_{2}+il$.
However, non-Hermitian quantum systems built upon two coupled units, one with gain and the other one with loss as shown in Fig.~\ref{fig:intro}(b), possess purely real eigenfrequencies in certain regimes as derived below. 
Note that such systems are also specifically termed parity-time-symmetric (PT-symmetric) systems~\cite{nn}.
The system dynamics in Fig.~\ref{fig:intro}(b) is expressed as
\begin{equation}
\label{eq:pt_the}
\dfrac{d}{dt} \left[\begin{array}{c}
 a_{\text{G}}\\
 a_{\text{L}}\\
\end{array}\right]=  \left[\begin{array}{cc}
i{\omega_1}+g & \kappa\\
\kappa & i{\omega_2}-l \\
\end{array}\right]\cdot \left[\begin{array}{c}
 a_{\text{G}}\\
 a_{\text{L}}\\
\end{array}\right],
\end{equation}
where the subscript G (or L) refers to the gain (or loss) unit and $a_{\text{G,L}}$ is the field amplitude defined such that $|a_{\text{G,L}}|^2$ represents the energy stored in each unit. 
$g$ (or $l$) presents the gain (or loss) of the unit. 
$\kappa$ indicates the coupling strength between the two units and $\omega_{1,2}$ represents the resonant frequency of each unit.
To find the eigenfrequencies, we let $a_{\text{G,L}}\propto{\exp^{i\omega t}}$ and obtain the characteristic equation as 
\begin{equation}
\big(i(\omega_1-\omega)+g\big)\cdot \big(i(\omega_2-\omega)-l \big)+{\kappa}^2=0.
\end{equation}
For a balanced system where the gain is equal to the loss, i.e., $g=l$, the solutions are then given by the following expression:
\begin{equation}
\label{eq:pt_solu}
\omega = \omega_0\pm\sqrt{{\kappa}^2-g^2}, ~\omega_0=(\omega_1+\omega_2)/2.
\end{equation}
Eq.~\eqref{eq:pt_solu} shows that when the coupling strength $\kappa$ is stronger than a threshold determined by the gain-loss contrast $g$, i.e., $\kappa > g=l$, the system has a pair of real eigenfrequencies.
Particularly, these eigenfrequencies could evolve with gain-loss contrast in a wide range as long as $g=l\in(0,\kappa)$.
This simple analysis suggests that the interplay between gain/loss and their coupling provides a new degree of tuning freedom, i.e., gain/loss tuning freedom, 
to modulate the behaviors of a system.
In the next section, we discuss how this physical principle (i.e., PT-symmetric topology) can be applied to design VCOs.

\begin{figure}[!t]
\centering
\includegraphics[width=1.0\linewidth]{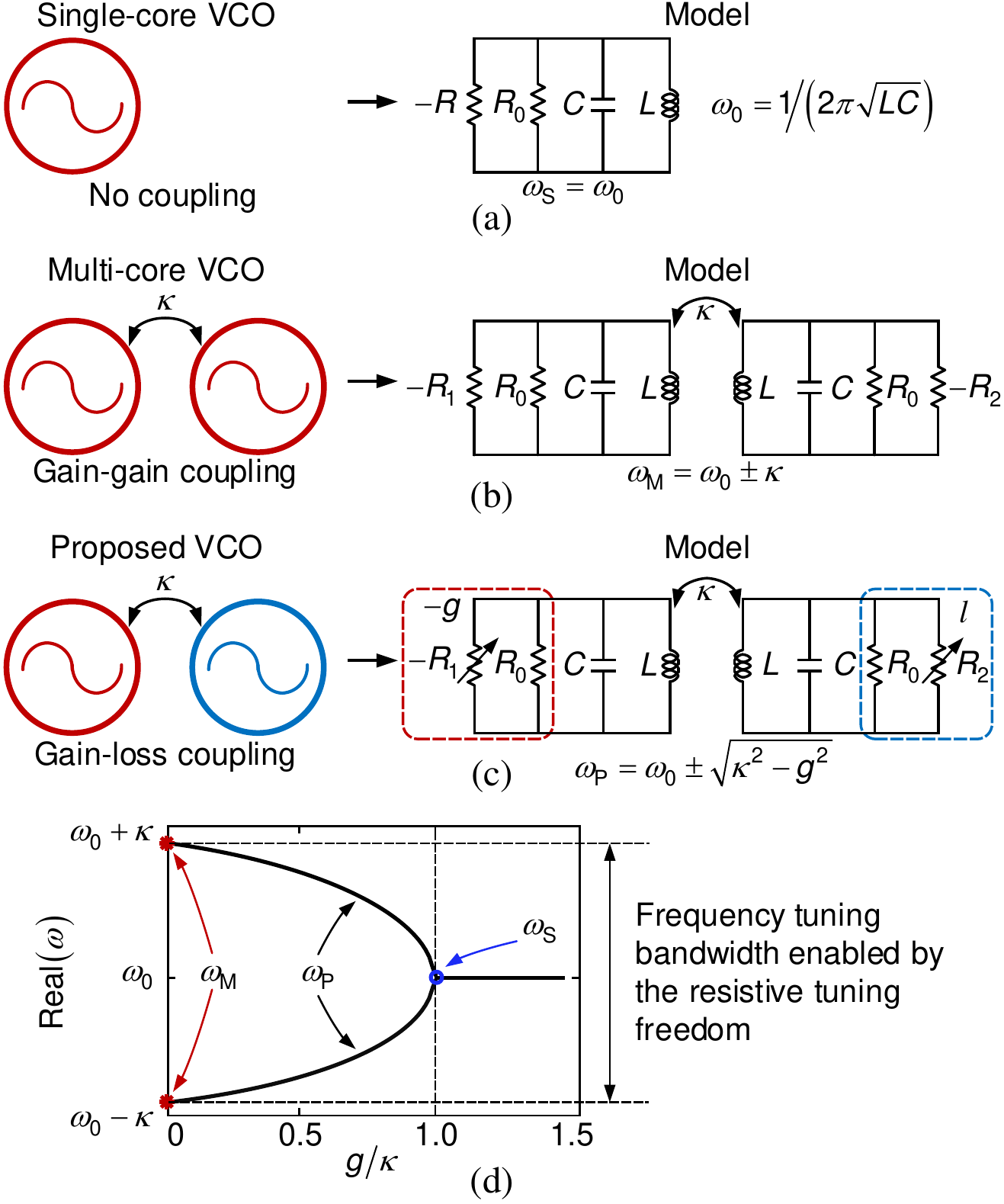}
\caption{Structure comparisons between (a) conventional single-core VCOs, (b) multi-core VCOs, and (c) proposed VCOs. (d) Numerical comparisons of frequency tuning between the three types of VCOs.}
\label{fig:con_vco}
\end{figure}


\subsection{Non-Hermitian Quantum Mechanics for VCO Design}

Before introducing the proposed VCO topology, Fig.~\ref{fig:con_vco} first re-examines conventional single-core and multi-core VCOs.
A single-core VCO can be simplified into an active LC resonator shown in Fig.~\ref{fig:con_vco}(a). 
It consists of a gain ($-R$) and an LC core with an intrinsic loss ($R_0$). 
A multi-core VCO (e.g., dual-core) is built upon two coupled single-core VCOs with a fixed coupling strength $\kappa$, each of which is simplified into an active LC resonator. 
For both VCOs, at the start-up phase, the gain is set to be slightly higher than the inherent loss to produce an oscillation with exponentially-growing amplitude. 
As the amplitude grows, the gain saturates and equates the loss in the large-signal domain due to nonlinearity. 
The oscillation then becomes stable, and the frequency (i.e., $\omega_{\text{S}} / \omega_{\text{M}}$ in Fig.~\ref{fig:con_vco}(a)/(b)) can only be adjusted via $\omega_0$ by tuning the core's capacitance or inductance. 
In particular, the multi-core VCO has two frequency modes as shown by $\omega_{\text{M}}$ in Fig.~\ref{fig:con_vco}(b).


This re-examination shows that in the conventional VCOs, the gain-loss distribution plays only a trivial role in the transient behavior of oscillators, i.e., the gain is used to compensate for the undesired loss to establish the start-up condition for the exponential growth of oscillation amplitude.
Fortunately, based on the non-Hermitian quantum mechanics introduced before, the loss is useful if the gain-loss distribution in a system is properly manipulated.
Inspired by this physical principle, our method explores the interplay between the gain/loss distribution and their coupling to enhance the frequency tuning bandwidth of VCOs. 
Fig.~\ref{fig:con_vco}(c) exhibits the simplified topology of the proposed VCO. 
It is built upon two coupled LC cores, one active with a negative resistance $-g$ and the other one dissipative with an equal amount of loss $l$, and the two cores have the same capacitance and inductance.
The proposed VCO exhibits two frequency modes as shown by $\omega_{\text{P}}$ in Fig.~\ref{fig:con_vco}(c) and an extra resistive tuning freedom (i.e., $g$) that is orthogonal with the typical capacitive/inductive tuning freedoms of $\omega_0$. 
Fig.~\ref{fig:con_vco}(d) numerically compares the frequency tuning of these three types of VCOs. 
Both the single-core VCO and multi-core VCO have only individual frequency points (i.e., blue circle and red asterisks), which are independent of $g$. 
However, the proposed VCO has a very wide FTR enabled by $g$.
The comparison shows that the proposed VCO with the resistive tuning freedom can achieve a wider FTR than conventional VCOs given the same capacitive/inductive tuning ability preserved by $\omega_0$.
\section{Gain-Loss Coupled Dual-Core VCO Topology}
\label{sec:imple}

\begin{figure}[!t]
\centering
\includegraphics[width=1.0\linewidth]{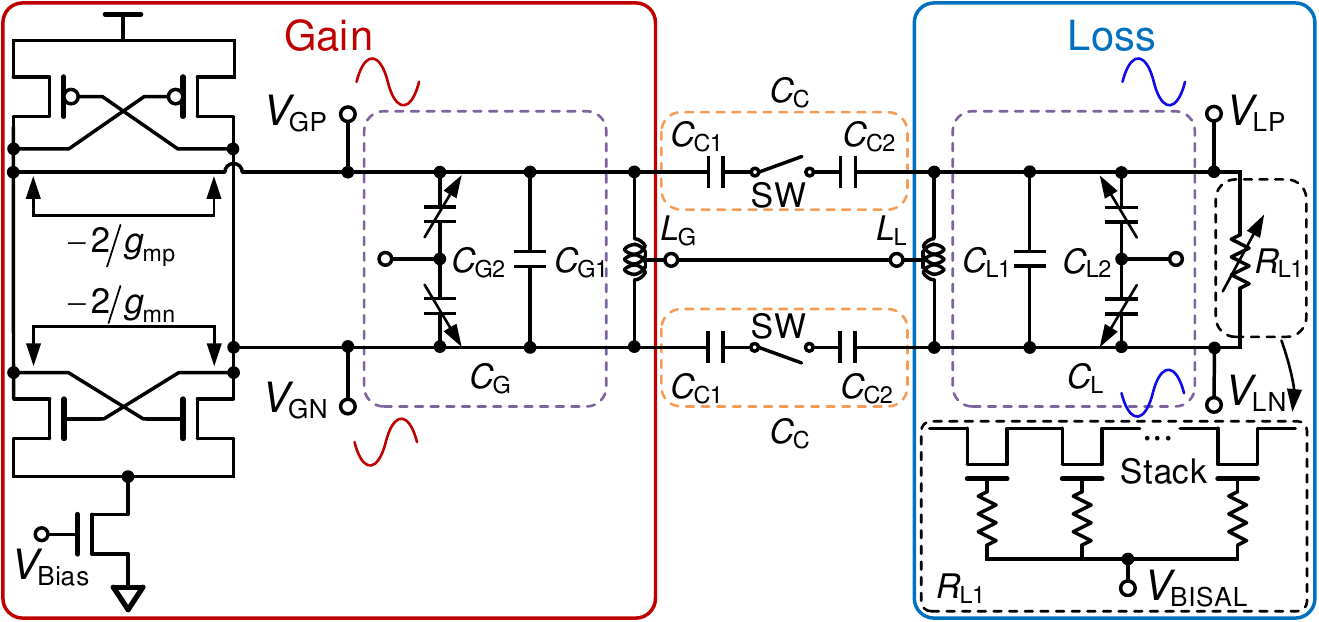}
\caption{The circuit schematic of the proposed VCO.}
\label{fig:circuit}
\end{figure}

\subsection{Circuit Design}
\label{sec:overall}

Fig.~\ref{fig:circuit} shows the proposed VCO circuit.
The gain side has a tunable gain rate generated by cross-coupled differential pairs (XDPs) and an inherent loss rate $R_{\text{G0}}$, leading to the total gain of $-R_{\text{G}}=(-1/G_m)\verb||||R_{\text{G0}}$;
$G_{m}= {(g^{}_{\text{mn}}+g^{}_{\text{mp}})/{2}}$, where $g^{}_{\text{mn}}$ and $g^{}_{\text{mp}}$ are the small signal transconductance of NMOS and PMOS XDP.
The loss side has an intrinsic loss rate $R_{\text{L0}}$ and a variable loss rate $R_{\text{L1}}$, giving the total loss of $R_{\text{L}}=R_{\text{L0}}\verb||||R_{\text{L1}}$.
To make loss adjustable, a variable resistor based on stacked transistors is parallelly connected to the loss side.
The subset in Fig.~\ref{fig:circuit} shows the schematic of the variable resistor $R_{\text{L1}}$.
All the gates of MOSFETs are connected together.
By tuning the gate voltage $V_{\text{BIASL}}$, the resistance can be continuously adjusted in a wide range.
The capacitor $C_{\text{G}}$ ($C_{\text{L}}$) in each LC core is composed of a parasitic capacitance $C_{\text{G0}}$ ($C_{\text{L0}}$), a fixed Metal-Insulator-Metal (MIM) capacitor $C_{\text{G1}}$ ($C_{\text{L1}}$) with high-quality factor (high-Q) and an adjustable varactor $C_{\text{G2}}$ ($C_{\text{L2}}$). 
The varactor takes up a small proportion of the total capacitance and is mainly used to compensate for the fabrication mismatch of the fixed MIM capacitors on each side. 
The coupling capacitance ($C_\text{C}$) is realized by two equal MIM capacitors ($C_{\text{C1}}$ and $C_{\text{C2}}$) which are serially connected through an on-chip switch (SW). 
The inductance ($L_\text{G}/L_\text{L}$) in both cores comes from the high-Q symmetrical parallel inductor (symindp) of the technology.
A center tap connection is provided such that both cores share the same common mode voltage by connecting the center taps of the inductors.
The balanced condition is satisfied by setting $R_{\text{G}} \approx R_{\text{L}}=R$, $L_{\text{G}} \approx L_{\text{L}}=L$, and $C_{\text{G}} \approx C_{\text{L}}=C$.

\subsection{Phase Noise Analysis}
\label{sec: pn_a}
We perform a qualitative analysis on the phase noise (PN) of the proposed VCO. 
It is well-established in the classic VCO theory that a multi-core VCO built upon a coupled structure can lead to PN reduction as compared to a single-core VCO.
By taking a two-core VCO as an example, the improved PN can be intuitively understood as that the equivalent current noise of each LC core experiences twice the capacitance, and therefore its PN contribution is reduced by $6$ dB. 
Two noise contributions from the two cores are uncorrelated and can be summed up, ideally leading to a $3$ dB reduction of the total PN. 
Generally, for an $N$-core VCO built upon a coupled structure, its PN is lower than a single-core VCO by $10\log_{10}N$ dB.
Thanks to the coupled structure,  the proposed VCO topology also inherits the PN advantage of conventional multi-core VCOs.
On the other hand, the gain-loss tuning physically realized by active devices, although it does not generate more inherent losses, 
does contribute additional noise to the system.
However, this is not a big issue as the unique gain-loss tuning also increases the carrier amplitude which suppresses the effect of noise.
It can be shown as follows.
For the proposed VCO, the gain not only compensates the inherent loss in the active core but also offsets the tunable loss in the coupled lossy core. 
Assuming the ratio between the tunable gain $-G_m$ generated by XDPs and the inherent loss $R_0$ is $\beta$ ($\beta>1$), the PN of the active core in our proposed VCO can be obtained based on the well-known PN model of single-core VCOs as below: 
\begin{equation}
\begin{split}
\label{eqn:pn_pt}
&\ \mathcal{L}_{\text{P}}(\triangle \omega) = \\
&\ 10  \log_{10} \Bigg((1+ \beta m)\cdot \frac{4kTR_0}{({\beta}^2 V_{\text{osc}})^2} \cdot\Big(\frac{\omega}{2Q_{S}\triangle \omega}\Big)^2\Bigg)\\
&\ = \underset{ \mathcal{L}_{\text{S}}(\triangle \omega)}{\underbrace{10  \log_{10} \Bigg((1 + m)\cdot \frac{4kTR_0}{(V_{\text{osc}})^2}
\cdot\Big(\frac{\omega}{2Q_{S}\triangle \omega}\Big)^2\Bigg)}}\\
&\ - \underset{>0}{\underbrace{10\log_{10}\Big(\dfrac{{\beta}^4(1+m)}{(1+\beta m)}\Big)}}<\mathcal{L}_{\text{S}}(\triangle \omega),
\end{split}
\end{equation}
where $k$ is the Boltzmann constant; $T$ is the absolute temperature; $R_0$ is the inherent resonator loss; $m$ ($m>1$) is a constant noise factor of active elements; $V_{\text{osc}}$ is the amplitude of the carrier; $Q_S$ is the quality factor of the LC core; $\mathcal{L}_{\text{S}}(\triangle \omega)$ is PN of a conventional single-core VCO.
Compared to the PN of conventional single-core VCOs, both the noise factor $m$ of active devices and the amplitude $V_{\text{osc}}$ of carrier increase in the PN formula of the active core of the proposed VCO. 
But the carrier amplitude increases to $\beta^2\times$ of the conventional one because the current flowing into the core is quadratically proportional to the gain when XDPs operate in the saturation region. 
Therefore, the proposed VCO topology shows better PN performance than conventional single-core VCOs.

\section{Experimental Evaluations}
\label{sec:results}

\begin{figure}[!t]
\centering
\includegraphics[width=0.55\linewidth]{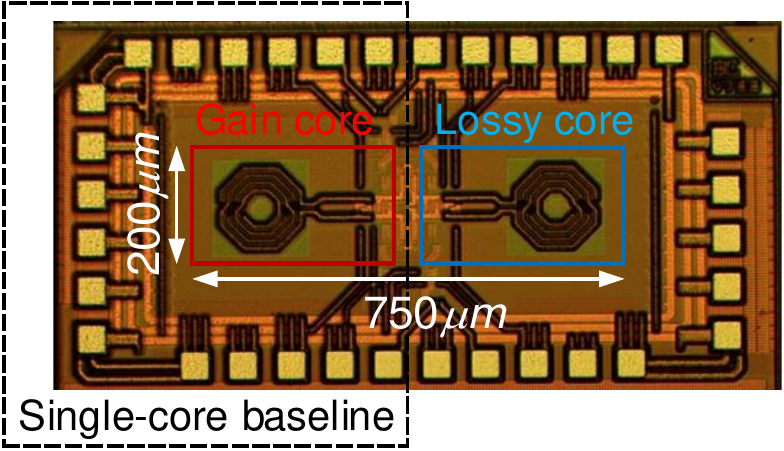}
\caption{Die micrograph of the proposed VCO.}
\label{fig:chip}
\end{figure}

To experimentally verify the advantages of the proposed VCO, a prototype design is implemented in a standard $130$ nm bulk CMOS process with a core area of $0.15$ $\text{mm}^2$ as shown in Fig.~\ref{fig:chip}. 
The two LC cores can be coupled (decoupled) by turning on (off) the switch SW. 
A single-core VCO, i.e., the active LC core in our design, is used as the baseline to directly compare with ours on the same monolithic chip. 
The baseline only has capacitive tuning freedom.
Additionally, since it inherently comes from our proposed VCO with the same non-ideal parasitic effects, thereby serving as a fair candidate for comparison to show the enhanced performance solely due to the contribution of the extra resistive tuning freedom.

\begin{figure}[!t]
\centering
\includegraphics[width=1.0\linewidth]{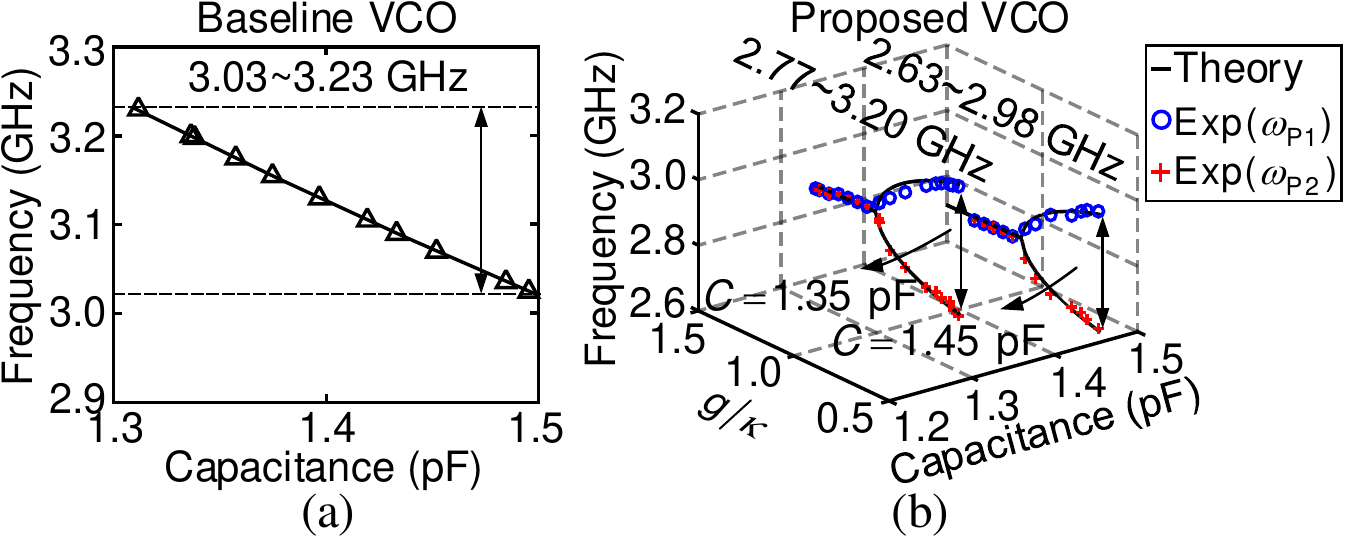}
\caption{Frequency tuning of the two VCOs. (a), The baseline VCO. (b), The proposed VCO. Theory: theoretical predictions; Exp: experimental results.}
\label{fig:ftr}
\end{figure}


\begin{table*}[!tb]
\footnotesize
\centering
\caption{Comparisons between the proposed VCO and state-of-the-art VCOs.}
\vspace{-0.1cm}
\begin{threeparttable}
\begin{tabular}{|c|c|c|c|c|c|c|c|c|c|c|c|c|c|}
\hline
\multicolumn{2}{|c|}{Reference}                                                         & \multicolumn{2}{c|}{JSSC '13~\cite{JSSC_13}}                              & \multicolumn{2}{c|}{TCAS-I '12~\cite{TCAS_I_12}}                                                              & \multicolumn{2}{c|}{ISSCC '19~\cite{ISSCC_19_s}}                                                                 & \multicolumn{2}{c|}{ISSCC '19~\cite{ISSCC_19}}                                                              & \multicolumn{2}{c|}{JSSC '17~\cite{JSSC_17}}                                                          & \multicolumn{2}{c|}{\text{This work}}                                              \\ \hline
\multicolumn{2}{|c|}{{VCO types}}                                        & \multicolumn{2}{c|}{Single-core}                   & \multicolumn{2}{c|}{Single-core}                                                   & \multicolumn{2}{c|}{Single-core}                                                         & \multicolumn{2}{c|}{Dual-core}                                                      & \multicolumn{2}{c|}{Quad-core}                                                  & \multicolumn{2}{c|}{Dual-core}                                               \\ \hline
\multicolumn{2}{|c|}{Optimization technique}                                                                  & \multicolumn{2}{c|}{\begin{tabular}[c]{@{}c@{}}Suppression of\\ flicker noise\end{tabular}}                           & \multicolumn{2}{c|}{\begin{tabular}[c]{@{}c@{}}Switched-coupled  inductor,\\ aggressive capacitive tuning \end{tabular}} & \multicolumn{2}{c|}{\begin{tabular}[c]{@{}c@{}}Narrowband   \\  resonance at $2f_\text{osc}$\end{tabular}} & \multicolumn{2}{c|}{\begin{tabular}[c]{@{}c@{}}Aggressive\\ capacitive tuning  \end{tabular}} & \multicolumn{2}{c|}{Multi-core} & \multicolumn{2}{c|}{\begin{tabular}[c]{@{}c@{}}\text{No}\\ \text{optimization}\end{tabular}} \\ \hline
\multicolumn{2}{|c|}{Technology}                                                           & \multicolumn{2}{c|}{$65$ nm CMOS}                    & \multicolumn{2}{c|}{$90$ nm CMOS}                                                    & \multicolumn{2}{c|}{$22$ nm FDSOI}                                                       & \multicolumn{2}{c|}{$65$ nm CMOS}                                                   & \multicolumn{2}{c|}{$55$ nm BiCMOS}                                                & \multicolumn{2}{c|}{\text{$130$ nm CMOS}}                                            \\ \hline
\multicolumn{2}{|c|}{Power supply (V)}                                                  & \multicolumn{2}{c|}{$1.2$}                           & \multicolumn{2}{c|}{$1.2$}                                                           & \multicolumn{2}{c|}{$0.15$}                                                              & \multicolumn{2}{c|}{$0.65$}                                                           & \multicolumn{2}{c|}{$1.2$}                                                       & \multicolumn{2}{c|}{\textbf{$1.2$}}                                                    \\ \hline

\multicolumn{2}{|c|}{Power (mW)}                                                   & \multicolumn{2}{c|}{$0.72$}                          & \multicolumn{2}{c|}{$1.06$}                                                          & \multicolumn{2}{c|}{$0.91\sim1.22$}                                                   & \multicolumn{2}{c|}{$17.5\sim21.6$}                                                 & \multicolumn{2}{c|}{$50$}                                 & \multicolumn{2}{c|}{\textbf{$2\sim4.31$}}                                                       \\ \hline
\multicolumn{2}{|c|}{Area ($\text{mm}^2)$}                                                 & \multicolumn{2}{c|}{$0.0806$}                              & \multicolumn{2}{c|}{$0.5$}                                                              & \multicolumn{2}{c|}{$0.272$}                                                       & \multicolumn{2}{c|}{$0.08$}                                                  & \multicolumn{2}{c|}{$0.6$}                                & \multicolumn{2}{c|}{\textbf{$0.15$}}                                                       \\ \hline

\multicolumn{2}{|c|}{{Tuning bandwidth (GHz)}}                                  & \multicolumn{2}{c|}{{$3.0\sim3.6$}} & \multicolumn{2}{c|}{{$1.13\sim1.9$}}                                & \multicolumn{2}{c|}{$4.15\sim4.97$}                                                   & \multicolumn{2}{c|}{$25\sim38$}                                                & \multicolumn{2}{c|}{$17.4\sim20.3$}                                              & \multicolumn{2}{c|}{\textbf{$2.63\sim3.20$}}                                            \\ \hline
\multicolumn{2}{|c|}{FTR (\%)}                                               & \multicolumn{2}{c|}{$18.2\%$}                    & \multicolumn{2}{c|}{$50.8\%$}                                                    & \multicolumn{2}{c|}{$18\%$}                                                       & \multicolumn{2}{c|}{$41.2\%$}                                                    & \multicolumn{2}{c|}{$15.3\%$}                                                & \multicolumn{2}{c|}{\textbf{$20.2\%$}}                                             \\ \hline
\multicolumn{2}{|c|}{Resistive tuning}                                                    & \multicolumn{2}{c|}{\xmark}                                                       &  \multicolumn{2}{c|}{\xmark}                                                                                    &  \multicolumn{2}{c|}{\xmark}                                            &\multicolumn{2}{c|}{\xmark}                                        &  \multicolumn{2}{c|}{\xmark}                                     & \multicolumn{2}{c|}{\textbf{\cmark}}                                   \\ \hline
\multicolumn{2}{|c|}{PN (dBc/Hz) (Average)}  & \multicolumn{2}{c|}{$-112$@$1$MHz}                     & \multicolumn{2}{c|}{$-117.2$@$1$MHz}                                      & \multicolumn{2}{c|}{$-141$@$10$MHz}                                    & \multicolumn{2}{c|}{$-116$@$3$MHz}                                & \multicolumn{2}{c|}{$-106.5$@$1$MHz}                                & \multicolumn{2}{c|}{\text{$-$$120.3$@$1$MHz}}                           \\ \hline
\multicolumn{2}{|c|}{FoM (dB)\tnote {a} ~(Average)}                                                          & \multicolumn{2}{c|}{$183$@$1$MHz}                      & \multicolumn{2}{c|}{$177.3$@$1$MHz}                                     & \multicolumn{2}{c|}{$193$@$10$MHz}                                      & \multicolumn{2}{c|}{$183$@$3$MHz}                                    & \multicolumn{2}{c|}{$187.5$@$1$MHz}                                  &\multicolumn{2}{c|}{\text{$184$@$1$MHz}}                                                                          \\ \hline
\multicolumn{2}{|c|}{$\text{FoM}_\text{T}$ (dB)\tnote {b} ~(Average)}                                                          & \multicolumn{2}{c|}{$188.2$@$1$MHz}                      & \multicolumn{2}{c|}{$191$@$1$MHz}                                     & \multicolumn{2}{c|}{$198$@$10$MHz}                                      & \multicolumn{2}{c|}{$195$@$3$MHz}                                    & \multicolumn{2}{c|}{$191$@$1$MHz}                                  &\multicolumn{2}{c|}{\text{$190.1$@$1$MHz}}                                                                          \\ \hline
\end{tabular}
\begin{tablenotes}\footnotesize
\item[a] $\text{FoM}=|{\text{PN}}|+20{\log}_{10}(f_{\text{osc}}/\triangle f)-10\log_{10}(P_{\text{DC}}/1\text{mW})$.
\item[b] $\text{FoM}_{\text{T}}=\text{FoM}+20\log_{10}({\text{FTR}}/10\%)$.
\end{tablenotes}
\end{threeparttable}
\vspace{-0.5cm}
\label{tb:com_1_comp}
\end{table*}
Fig.~\ref{fig:ftr} shows the frequency tuning curves of the two VCOs. 
The baseline yielded a $0.20$ GHz ($3.03\sim3.23$ {GH}z, $6.4\%$ FTR) bandwidth tuning as demonstrated in Fig.~\ref{fig:ftr}(a) by adjusting the control voltage of varactors.
Such a tuning range corresponds to a capacitive tuning ability of $[1.30,1.50]$ pF.
We then set the core capacitance to be $1.35$ {pF} and $1.45$ {pF} respectively.
Fig.~\ref{fig:ftr}(b) exhibits the tuning curves corresponding to each capacitance value.
At $C=1.35$ {pF} ($C=1.45$ {pF}), the proposed VCO achieves a tuning bandwidth of $[2.77, 3.20]$ GHz ($[2.63, 2.98]$ GHz) with the extra resistive tuning freedom.
The results show that even with a slightly reduced amount of the capacitive tuning ability, the proposed VCO can realize a wider bandwidth tuning of $0.57$ {GH}z ($2.63\sim3.20$ {GH}z, $20.2\%$ FTR) by including the resistive tuning freedom, enabling a $3.1 \times$ FTR of the baseline.
Note that this prototype only a small range of capacitive tuning ability (i.e., $[1.30, 1.50]$ pF) is included in this prototype, thereby achieving an FTR of $20\%$. 
By slightly increasing the capacitive tuning ability, the proposed VCO can readily realize a wider FTR. 

\begin{figure}[!t]
\centering
\includegraphics[width=1.0\linewidth]{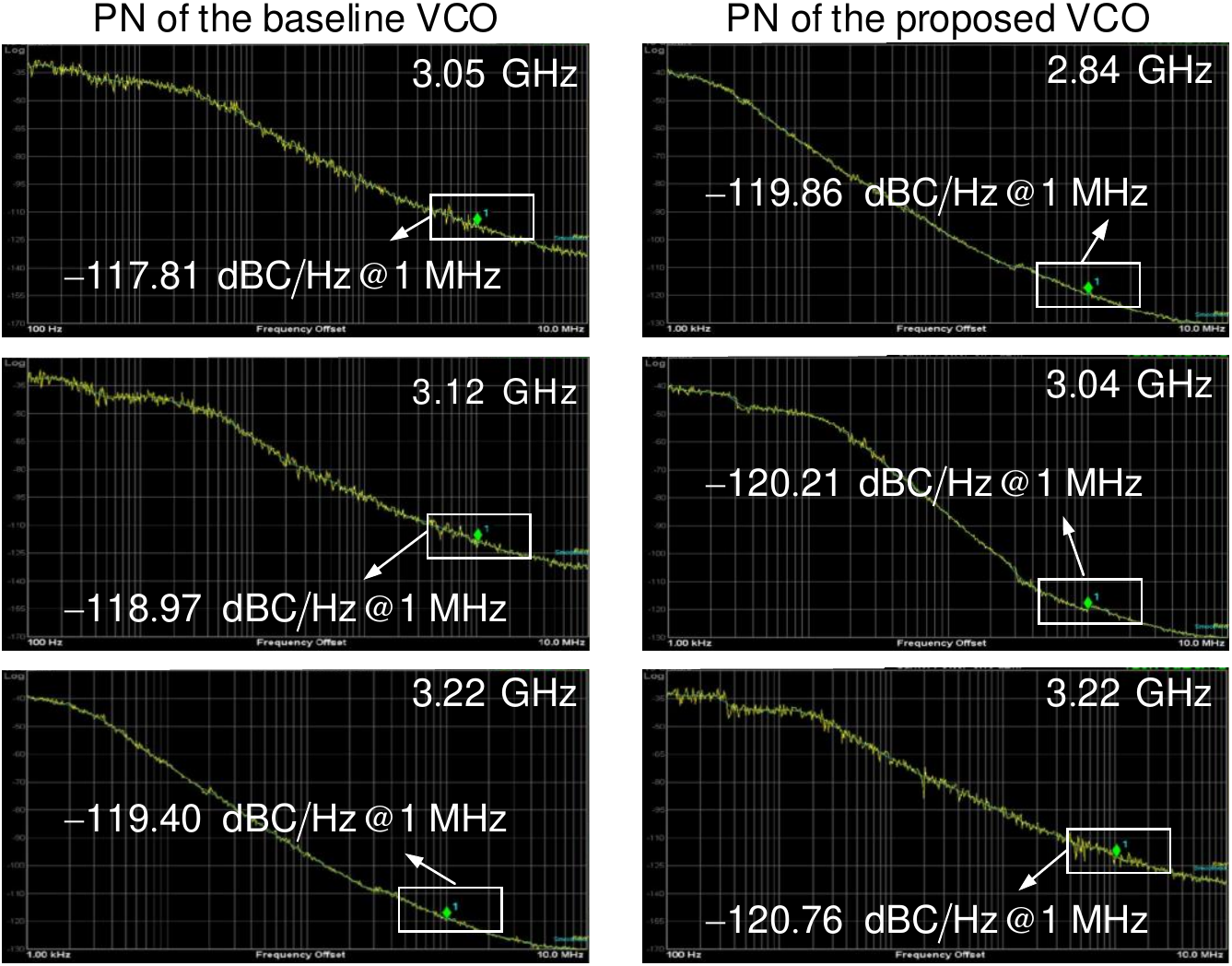}
\vskip -3pt
\caption{PN comparisons of two VCOs across different oscillation frequencies.}
\label{fig:pn}
\vskip -9pt
\end{figure}



We then show the PN of both VCOs.
Particularly, we measure the PN at three frequency points for each VCO in its tuning bandwidth.
For the proposed VCO, we choose such frequencies to be $2.84$ GHz (low), $3.04$ GHz (medium), and $3.22$ GHz (high).
While for the baseline VCO, we choose them to be $3.05$ GHz (low), $3.12$ GHz (medium), and $3.22$ GHz (high).
Fig.~\ref{fig:pn} shows the measured PN at $1$ MHz offset frequency for the two VCOs.
We observed that the PN of the proposed VCO is generally $1.5$ dB better than the baseline across the different oscillation frequencies.
Due to the parasitics of the switch connecting the two LC cores, the PN improvement is not as much as the ideal case discussed in Section~\ref{sec: pn_a}.
However, these observations generally match well with the previous qualitative characterizations.
Our results show that manipulating the gain-loss profile and their coupling provides a new method to extend the frequency tuning dimension beyond conventional capacitive/inductive tuning without compromising the PN performance for VCO design. 



Additionally, we compare the proposed VCO with other conventional VCOs, i.e., single-core VCOs, two-core VCOs, and quad-core VCOs that employ different tuning manners or coupling structures, as summarized in Table~\ref{tb:com_1_comp}.
These conventional VCOs exploit different optimization techniques, such as aggressive capacitive tuning to reach wide FTR (TCAS-I '12~\cite{TCAS_I_12}), narrow band resonance at $2f_{\text{osc}}$ to boost PN performance (ISSCC '19~\cite{ISSCC_19_s})), and multi cores to enhance PN performance (JSSC '17~\cite{JSSC_17}).
However, our proposed VCO only includes a resistive tuning into design without any other optimization techniques.
The comparison still shows its 
comparable FTR, PN, and figure-of-merit (FoM)  with these prior arts.
In summary, the resistive tuning of our proposed VCO topology is orthogonal with other capacitive/inductive tuning dimensions to enhance the performance of VCOs with diverse conventional topologies and optimization techniques.

\section{Conclusion}
\label{sec:conclu}
A non-Hermitian physics-inspired topology of VCO is shown in this paper.
The new topology enhances the FTR of existing VCOs with extra resistive tuning freedom.
A prototype is implemented in a standard $130$ nm CMOS process to demonstrate its advantages.
The comparisons show that the resistive tuning of our proposed VCO topology is orthogonal with other capacitive/inductive tuning dimensions to enhance the performance of VCOs with diverse conventional topologies and optimization techniques.
Future explorations can be performed by cascading multiple such VCOs in a one-dimensional chain similar to the one shown in prior work~\cite{ns}, which may also achieve topological oscillators.

\bibliographystyle{IEEEtran}

\bibliography{main,main}

\begin{thebibliography}{10}
\providecommand{\url}[1]{#1}
\csname url@samestyle\endcsname
\providecommand{\newblock}{\relax}
\providecommand{\bibinfo}[2]{#2}
\providecommand{\BIBentrySTDinterwordspacing}{\spaceskip=0pt\relax}
\providecommand{\BIBentryALTinterwordstretchfactor}{4}
\providecommand{\BIBentryALTinterwordspacing}{\spaceskip=\fontdimen2\font plus
\BIBentryALTinterwordstretchfactor\fontdimen3\font minus
  \fontdimen4\font\relax}
\providecommand{\BIBforeignlanguage}[2]{{%
\expandafter\ifx\csname l@#1\endcsname\relax
\typeout{** WARNING: IEEEtran.bst: No hyphenation pattern has been}%
\typeout{** loaded for the language `#1'. Using the pattern for}%
\typeout{** the default language instead.}%
\else
\language=\csname l@#1\endcsname
\fi
#2}}
\providecommand{\BIBdecl}{\relax}
\BIBdecl

\bibitem{fangxu}
F.~Lv, X.~Zheng, F.~Zhao, J.~Wang, S.~Yue, Z.~Wang, W.~Cao, Y.~He, C.~Zhang,
  H.~Jiang, and Z.~Wang, ``{A power scalable 2-10 Gb/s PI-based clock data
  recovery for multilane applications},'' \emph{Microelectronics Journal},
  vol.~82, pp. 36--45, 2018.

\bibitem{naiwen}
N.~Zhou, L.~Wu, Z.~Wang, X.~Zheng, W.~Cao, C.~Zhang, F.~Li, and Z.~Wang, ``{A
  28-Gb/s transmitter with 3-tap FFE and T-coil enhanced terminal in 65-nm CMOS
  technology},'' in \emph{2016 14th IEEE International New Circuits and Systems
  Conference (NEWCAS)}, 2016, pp. 1--4.

\bibitem{caodfe}
W.~Cao, Z.~Wang, D.~Li, X.~Zheng, K.~Huang, S.~Yuan, F.~Li, and Z.~Wang, ``{A
  40Gb/s 27mW 3-tap closed-loop decision feedback equalizer in 65nm CMOS},'' in
  \emph{2015 IEEE 13th International New Circuits and Systems Conference
  (NEWCAS)}.\hskip 1em plus 0.5em minus 0.4em\relax IEEE, 2015, pp. 1--4.

\bibitem{TCAS_I_12}
A.~I. et~al, ``{A 1-mW 1.13–1.9 GHz CMOS LC VCO Using Shunt-Connected
  Switched-Coupled Inductors},'' \emph{IEEE Transactions on Circuits and
  Systems I: Regular Papers}, vol.~59, no.~6, pp. 1145--1155, 2012.

\bibitem{VLSI_10_IND}
{Tanabe, Akira et al}, ``{A 5–20GHz tunable LC-VCO using variable bridge
  inductor},'' in \emph{2010 Symposium on VLSI Circuits}, 2010, pp. 47--48.

\bibitem{JSSC_13}
F.~Pepe, A.~Bonfanti, S.~Levantino, C.~Samori, and A.~L. Lacaita,
  ``{Suppression of Flicker Noise Up-Conversion in a 65-nm CMOS VCO in the
  3.0-to-3.6 GHz Band},'' \emph{IEEE Journal of Solid-State Circuits}, vol.~48,
  no.~10, pp. 2375--2389, 2013.

\bibitem{ISSCC_19}
A.~Bhat and N.~Krishnapura, ``{26.3 A 25-to-38GHz, 195dB $\text{FoM}_\text{T}$
  LC QVCO in 65nm LP CMOS Using a 4-Port Dual-Mode Resonator for 5G Radios},''
  in \emph{2019 IEEE International Solid- State Circuits Conference - (ISSCC)},
  2019, pp. 412--414.

\bibitem{JSSC_17}
L.~Iotti, A.~Mazzanti, and F.~Svelto, ``{Insights Into Phase-Noise Scaling in
  Switch-Coupled Multi-Core LC VCOs for E-Band Adaptive Modulation Links},''
  \emph{IEEE Journal of Solid-State Circuits}, vol.~52, no.~7, pp. 1703--1718,
  2017.

\bibitem{hai_kun}
W.~Deng, H.~Jia, R.~Wu, S.~Sun, C.~Li, Z.~Wang, and B.~Chi, ``An 8.2-to-21.5
  ghz dual-core quad-mode orthogonal-coupled vco with concurrently dual-output
  using parallel 8-shaped resonator,'' in \emph{2021 IEEE Custom Integrated
  Circuits Conference (CICC)}, 2021, pp. 1--2.

\bibitem{nn}
W.~Cao, C.~Wang, W.~Chen, S.~Hu, H.~Wang, L.~Yang, and X.~Zhang, ``{Fully
  integrated parity--time-symmetric electronics},'' \emph{Nature
  nanotechnology}, vol.~17, no.~3, pp. 262--268, 2022.

\bibitem{NM}
{\c{S}}.~K. {\"O}zdemir, S.~Rotter, F.~Nori, and L.~Yang, ``Parity--time
  symmetry and exceptional points in photonics,'' \emph{Nature Materials},
  vol.~18, no.~8, pp. 783--798, Aug 2019.

\bibitem{ISSCC_19_s}
O.~El-Aassar and G.~M. Rebeiz, ``{26.5 A 0.1-to-0.2V Transformer-Based
  Switched-Mode Folded DCO in 22nm FDSOI With Active Step-Down Impedance
  Achieving 197dBc/Hz Peak FoM and 40MHz/V Frequency Pushing},'' in \emph{2019
  IEEE International Solid- State Circuits Conference - (ISSCC)}, 2019, pp.
  416--418.

\bibitem{ns}
Y.~Liu, W.~Cao, W.~Chen, H.~Wang, L.~Yang, and X.~Zhang, ``{Fully integrated
  topological electronics},'' \emph{Scientific reports}, vol.~12, no.~1, p.
  13410, 2022.

\end{thebibliography}


\begin{thebibliography}{00}
\bibitem{b1} G. Eason, B. Noble, and I. N. Sneddon, ``On certain integrals of Lipschitz-Hankel type involving products of Bessel functions,'' Phil. Trans. Roy. Soc. London, vol. A247, pp. 529--551, April 1955.
\bibitem{b2} J. Clerk Maxwell, A Treatise on Electricity and Magnetism, 3rd ed., vol. 2. Oxford: Clarendon, 1892, pp.68--73.
\bibitem{b3} I. S. Jacobs and C. P. Bean, ``Fine particles, thin films and exchange anisotropy,'' in Magnetism, vol. III, G. T. Rado and H. Suhl, Eds. New York: Academic, 1963, pp. 271--350.
\bibitem{b4} K. Elissa, ``Title of paper if known,'' unpublished.
\bibitem{b5} R. Nicole, ``Title of paper with only first word capitalized,'' J. Name Stand. Abbrev., in press.
\bibitem{b6} Y. Yorozu, M. Hirano, K. Oka, and Y. Tagawa, ``Electron spectroscopy studies on magneto-optical media and plastic substrate interface,'' IEEE Transl. J. Magn. Japan, vol. 2, pp. 740--741, August 1987 [Digests 9th Annual Conf. Magnetics Japan, p. 301, 1982].
\bibitem{b7} M. Young, The Technical Writer's Handbook. Mill Valley, CA: University Science, 1989.
\end{thebibliography}

\end{document}